# Marginally stable solutions

Pierre van Baal [a] *

[a]Instituut-Lorentz for Theoretical Physics,
University of Leiden, PO Box 9506,
NL-2300 RA Leiden, The Netherlands.

In previous work constant magnetic field strength solutions for SU(2) gauge theory on a torus were found, which somewhat surprisingly turned out to be classically stable. This was called marginal stability, as moving along one of its zero-modes, two of the stable modes turn unstable. Here we investigate the stability under quantum fluctuations in the domain where the solutions possess the marginal stability at the classical level.

Constant curvature solutions in gauge theories have frequently been studied for non-Abelian gauge theories, where it was realized that a constant magnetic field is an unstable solution of the equations of motion, due to the self-coupling of the gauge fields [1]. Attempts were made to find stable configurations that evolved from these constant magnetic fields and was the basis for the Copenhagen vacuum picture [2]. A regular two-dimensional lattice of flux configurations was found, whose periodicity turned out to be equivalent to twisted boundary conditions [3], which were formulated by 't Hooft in a different context [4]. Compactifying in all spatial directions leads to an infrared cutoff and renders a perturbative calculation reliable in a small volume [5].

In the four dimensional context it can be shown that a constant field is stable if it is self-dual [6,7], and in the course of the stability analysis for non-selfdual configurations the full fluctuation spectrum was determined [8]. The analysis in three dimensions is obtained by adding a periodic fourth direction, and leads to a degenerate twist. This, in the four-dimensional context exceptional, but in three dimensions generic case, was not studied fully. Thus we missed an amusing and surprising result, which was discovered while designing and testing cooling algorithms for saddle-points that are not local minima of the energy (or action) functional [9]. The most general constant curvature solution in SU(2) gauge theories *on a torus* has as many gauge invariant parameters (moduli) as the number of dimensions. By considering the gauge invariant Polyakov-loop observables it is easily seen that the constant curvature solution, compactified to a torus, is not translation invariant. In the generic four-dimensional case the moduli are equivalent to translations and from this it is obvious that the fluctuation spectrum is independent of the moduli. However, if the twist is degenerate on a $d$ dimensional subspace, the vector potential is invariant under translations in this subspace, and the Polyakov loops label moduli on which the fluctuation spectrum can and does depend. For three dimensions this is the generic case. Restricting to the case of non-zero field only, the subspace along which the solution has translation invariance is one dimensional.

For one class of constant magnetic field solutions on a torus (corresponding to the smallest possible non-zero magnetic field), it turns out that the usually unstable modes depend on the moduli in such a way that they turn stable for a limited range of values. Moving along the moduli (which does not change the energy) in the stable region, will bring one to the region where two fluctuation modes will become unstable, and the configuration decays. The effects we study here are absent in a (naive) infinite volume limit, where infrared divergencies, however, render perturbative arguments ambiguous. Unfortunately, we will find that quantum fluctuations will render the marginally stable configurations unstable. As such, this observation is only of marginal in-

---

terest. Note that for weak coupling the time scale involved in the decay is much longer than for the decay of classically unstable solutions.

## 1. All constant magnetic field solutions

For SU(2) gauge theories it can be shown [6] that solutions of the equations of motion with constant field strength are given, up to a gauge transformation, by

$$A_\mu^o = -iF_{\mu\nu}x_\nu \tau_3/2, \tag{1}$$

where $F_{\mu\nu}$ is a constant anti-symmetric tensor. On a torus, in addition we have to require that the vector potential is periodic up to a gauge transformation [4]. For convenience only we will restrict our attention to a symmetric torus of unit length. So for suitable gauge functions $\Omega_\nu(x) \in$ SU(2)

$$A_\mu^o(x+\hat{\nu}) = \Omega_\nu(x)(A_\mu^o(x)+\partial_\mu)\Omega_\nu^{-1}(x). \tag{2}$$

This can be read as a differential equation for these gauge functions, which can be cast in a simple form by writing

$$\Omega_\nu(x) = \exp(A_\nu^o(x))\omega_\nu(x). \tag{3}$$

Without further assumptions on $\omega_\nu(x)$ this yields

$$D_\mu \omega_\nu(x) = \partial_\mu \omega_\nu(x) + [A_\mu^o(x), \omega_\nu(x)] = 0, \tag{4}$$

for any $\mu$ and $\nu$. It is now straightforward to show (see [8], app. B) that $\omega_\nu(x)$ are constant and commute with $\tau_3$. They can be parametrized by constants $C_\nu$ mod $2\pi$ as

$$\omega_\nu(x) = \exp(iC_\nu \tau_3/2), \tag{5}$$

Multiplying $\omega_\nu(x)$ with a non-trivial element of the center of the gauge group ($-1$ for SU(2)), does not effect the boundary conditions of the gauge fields (not only for $A_\mu^o(x)$, but for *all* of them).

To disentangle the gauge invariant parameters it is useful to consider the Polyakov loops [9,10]

$$\begin{aligned} P_\mu(x) &= \tfrac{1}{2}\mathrm{Tr}\left[\mathrm{Pexp}\left(\int_0^1 A_\mu^o(x)dx_\mu\right)\Omega_\mu(x)\right] \\ &= \cos(\tfrac{1}{2}C_\mu - F_{\mu\nu}x_\nu), \end{aligned} \tag{6}$$

which shows that $C_\mu$ are gauge invariant parameters. By performing an overall gauge transformation we can transform $\omega_\nu(x)$ to the identity, and shift $C_\nu$ to the vector potential

$$A_\mu^C(x) = -iF_{\mu\nu}x_\nu\tau_3/2 + iC_\mu\tau_3/2. \tag{7}$$

Anti-periodic gauge transformations leave the boundary conditions invariant (but flip the sign of $\Omega_\nu(x)$ and $P_j(x)$). They shift the values of $C_\mu$ by $2\pi$. Also we need to require that $|P_j(x)|$ is a periodic function in all directions (the sign is again associated to the center of the gauge group and at the heart of the twisted boundary conditions [4]). This is seen to imply

$$F_{\mu\nu} = \pi n_{\mu\nu}, \tag{8}$$

with $n_{\mu\nu}$ integer and $Z_{\mu\nu} = \exp(-iF_{\mu\nu})$ are precisely the twist factors [4], which are the extra factors that arise when interchanging $\mu$ and $\nu$ in $\Omega_\mu(x+\hat{\nu})\Omega_\nu(x)$.

We note that if $F_{\mu\nu}$ is an invertible matrix the $C_\mu$ can be absorbed by a translation

$$A_\mu^C(x) = -iF_{\mu\nu}(x_\nu - y_\nu)\tau_3/2, \quad C_\mu = F_{\mu\nu}y_\nu. \tag{9}$$

However, in three dimensions any non-zero anti-symmetric tensor has precisely one degenerate eigenvector which can be identified with the magnetic flux vector defined by $m_i = \tfrac{1}{2}\varepsilon_{ijk}n_{jk}$ and obviously $P_j(\vec{x}+s\vec{m}) = P_j(\vec{x})$. We hence identify the moduli

$$c \equiv \vec{C}\cdot \vec{m}, \tag{10}$$

which is not related to a translation. Note that translation in the direction $\vec{m}$ leaves $A_\mu^o(x)$ invariant such that the number of moduli is equal to the number of dimensions. Obviously, the spectrum of fluctuations around the constant magnetic field is not effected by translations, but it can (and will, as we know form our experience for $F_{\mu\nu} = 0$) depend on the parameter $c$. It is also useful to note that any gauge field that satisfies the boundary conditions of eq. (2) is periodic in the direction $\vec{m}$. This is most easily seen by repeatedly applying eq. (2). Using the explicit abelian form of $\Omega_k(x)$ (see eqs. (3,5)) one finds

$$A_j(\vec{x}+\vec{m}/e) = \tau_3^{\Delta(\vec{m})} A_j(\vec{x}) \tau_3^{-\Delta(\vec{m})}, \tag{11}$$

where $e$ is the greatest common divisor of the $m_i$ and $\Delta(\vec{m}) = m_1 m_2 m_3/e^2$. One easily verifies that this implies periodicity, except if $\vec{m} = (1,1,1)$ mod 2. In that case the $\tau_3$ component of $\vec{A}$ is periodic, whereas the other two are anti-periodic. Here we will only be concerned with the marginally stable case, for which there is a non-empty interval in $c$ with *all* fluctuations stable. This is realized only [9] by $\vec{m} = (0,0,1)$ (or any $\vec{m}$ equivalent to it under the cubic symmetry group).

## 2. The full fluctuation spectrum

We compute here for $\vec{m} = (0,0,1)$ the transverse fluctuation spectrum on a symmetric torus of unit length. We keep the moduli not related to a translation in the expression for the vector potential and absorb the translation parameters in the boundary conditions

$$\vec{A}^c(\vec{x}) = \vec{A}^o(\vec{x}) + ic\vec{m}\tau_3/2,$$
$$\Omega_k(\vec{x}) = \exp(i\pi\vec{m}\wedge(\vec{x}-\vec{y})\tau_3/2). \quad (12)$$

Expanding the vector potential in components as

$$A_j = A_j^c - \frac{i}{\sqrt{2}}d_j\tau_+ - \frac{i}{\sqrt{2}}d_j^*\tau_- - \frac{i}{2}b_j\tau_3, \quad (13)$$

one finds the following boundary conditions

$$b_j(\vec{x}+\hat{k}) = b_j(\vec{x}), \quad (14)$$
$$d_j(\vec{x}+\hat{k}) = \exp\left(i\pi\hat{k}\cdot[\vec{m}\wedge(\vec{x}-\vec{y})]\right)d_j(\vec{x}).$$

The fluctuation operator splits as follows [8]

$$d_z^*(M-4\pi)d_z + d_{\bar{z}}^*(M+4\pi)d_{\bar{z}} + d_3^*Md_3 - \tfrac{1}{2}b_k\partial_j^2 b_k \quad (15)$$

with

$$z = \frac{x_1 - ix_2}{\sqrt{2}}, \quad d_z = \frac{d_1 + id_2}{\sqrt{2}}, \quad d_{\bar{z}} = \frac{d_1 - id_2}{\sqrt{2}} \quad (16)$$

and $M = -D_j^2$, where $D_j$ is the covariant derivative with respect to $A^c$, which on the $b_j$ modes reduces to the ordinary derivative, but on the $d_j$ modes leads to

$$M = aa^\dagger + a^\dagger a + (i\partial_3 - c)^2, \quad a = -i\partial_{\bar{z}} - i\pi z, \quad (17)$$

where $a$ can be seen as an annihilation operator, $[a, a^\dagger] = 2\pi$. Also $[M, a] = -4\pi a$, $[M, a^\dagger] = 4\pi a^\dagger$ and $[\partial_3, M] = 0$. From this we find that the eigenvalues of the fluctuation operator are given by

$$\lambda_{n,k}(c) = 2\pi(2n-1) + (2\pi k - c)^2, \quad n\in\mathbb{N}, k\in\mathbb{Z} \quad (18)$$

for the $d_j$ and $\lambda_{\vec{k}} = (2\pi\vec{k})^2$, $\vec{k}\in\mathbb{Z}^3$ for the $b_j$ modes. Transversality for the latter is imposed by demanding $\partial_j b_j = 0$ and are hence two-fold degenerate. To find the other degeneracies we consider the eigenfunctions of the $d_j$ modes in terms of the eigenfunctions of $M$

$$\Psi_{n,k}(\vec{x}) = (a^\dagger)^n \chi_o(z,\bar{z})e^{2\pi ikx_3}/\sqrt{n!}, \quad (19)$$

such that the eigenfunctions for the eigenvalue of eq. (18) are given by $d_z = \Psi_{n,k}$, $d_3 = \Psi_{n-1,k}$ ($n\neq 0$) and $d_{\bar{z}} = \Psi_{n-2,k}$ ($n\neq 0,1$). Transversality,

$$-iD_j d_j \tau_+ = [ad_z + a^\dagger d_{\bar{z}} - (c+i\partial_3)d_3]\tau_+ = 0, \quad (20)$$

is seen to give one linear constraint and leaves a four-fold degeneracy (as $d_j$ is a complex mode) for $n > 1$. For $n = 1$ there is no $d_{\bar{z}}$ eigenfunction and we have a two-fold degeneracy, whereas for $n = 0$ only $d_z$ survives as an eigenfunction, which is by itself transverse, since $\chi_0(z,\bar{z})$ is annihilated by $a$. One easily verifies [8] that

$$\chi_o(z,\bar{z}) = e^{-\pi\bar{z}z}f_o(z-y), \quad (21)$$

with $y = (y_1 - iy_2)/\sqrt{2}$ and $f_o(z)$ a holomorphic function satisfying the boundary conditions implied by eq. (14). For $\vec{m} = (0,0,1)$ there is only one such function, related to a Riemann theta function (for the explicit form see ref. [8], eq. (3.24)).

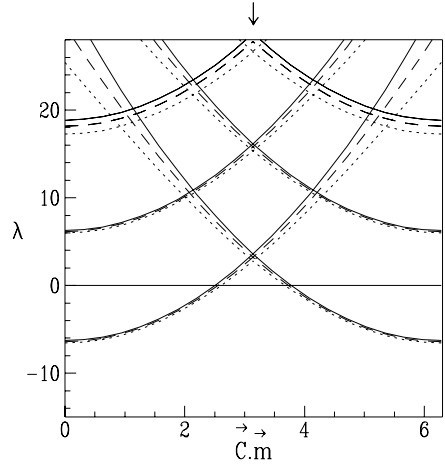

Figure 1. Lowest eigenvalues for the fluctuation operator for the continuum (full lines), a $4^3$ (dotted lines) and $6^3$ (dashed lines) lattices. The levels indicated by the arrow are four-fold and the others are two-fold degenerate. From ref. [9]

Except for $n = 0$ all eigenvalues are positive. Two negative eigenvalues appear for $|c - 2\pi k| < \sqrt{2\pi}$ in $\lambda_{0,k}(c)$. The region of stability for $c \in [0, 2\pi]$ is therefore given by $\sqrt{2\pi} < c < 2\pi - \sqrt{2\pi}$. In figure 1 we present the spectrum for $n = 0, 1, 2$ and $k = 0, 1$. The continuum result (full lines) are compared with numerical results on a $4^3$ (dotted lines) and $6^3$ (dashed lines) lattice for the fluctuation spectrum around the simple transcription of $\vec{A}^c(\vec{x})$ to a lattice, which is also an exact solution of the lattice equations of motion [9]. Note that the region of stability is reduced with decreasing lattice spacing, but persists.

## 3. The effective potential

Having identified the eigenvalues and degeneracies, it is straightforward to compute the effective potential in the region of stability.

$$V_{\text{eff}}(c) = 2 \sum_{n \in \mathbb{N}, k \in \mathbb{Z}} [\lambda_{n,k}^{\frac{1}{2}}(c) + \lambda_{n+2,k}^{\frac{1}{2}}(c)]. \qquad (22)$$

Using the heat kernel expansion gives

$$V_{\text{eff}}(c) = \frac{2}{\Gamma(-\frac{1}{2})} \sum_{n \in \mathbb{N}, k \in \mathbb{Z}} \int_0^\infty dt \, \frac{1 + e^{-8\pi t}}{t^{3/2}} e^{-t\lambda_{n,k}(c)}$$

$$= -\pi^{-\frac{1}{2}} \sum_{k \in \mathbb{Z}} \int_0^\infty dt \, \frac{(1 + e^{-8\pi t}) e^{-t\lambda_{0,k}(c)}}{(1 - e^{-4\pi t}) t^{3/2}}. \quad (23)$$

We split the integration over $t$ in the intervals $[0, 1]$ and $[1, \infty]$. For the integration over the first interval, Poisson resummation is applied

$$\sum_{k \in \mathbb{Z}} e^{-t(2\pi k - c)^2} = (4\pi t)^{-\frac{1}{2}} \sum_{\ell \in \mathbb{Z}} \cos(\ell c) \, e^{-\ell^2/4t}. \quad (24)$$

This leaves only the integral for the $\ell = 0$ term divergent, which can be dropped, since it is independent of $c$. All other integrals are finite, even at the edge of the stability region and can be performed numerically with high accuracy. The remaining sum is rapidly converging. The following expression is accurate to better than $1 : 10^5$

$$V_{\text{eff}}(c) = 2\lambda_{0,0}^{\frac{1}{2}}(c) + 2\lambda_{0,1}^{\frac{1}{2}}(c) + \alpha(c-\pi)^2 + \beta(c-\pi)^4, (25)$$

with $\alpha = -0.517943$ and $\beta = 0.011789$, which are determined from the exact Taylor coefficients. In figure 2 we plot $V_{\text{eff}}(c)$, normalized to vanish at the endpoints. It is obvious that the quantum fluctuations render these solutions unstable.

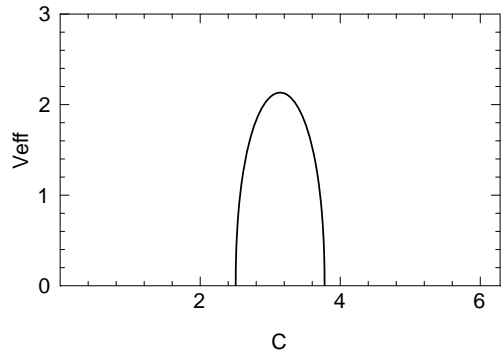

Figure 2. Effective potential in the stable region.

I thank Margarita García Pérez for a fruitful collaboration, studying instantons and sphalerons by cooling. I also thank Conrad Burden and Jim McGarthy for hospitality in Canberra and Adelaide. This work was supported in part by a grant from "Stichting Nationale Computer Faciliteiten (NCF)" for use of the CRAY C98 at SARA.